\documentclass[prd,aps,floats,preprint,preprintnumbers]{revtex4}
\usepackage{epsfig}
\usepackage{amsmath} 
\usepackage{amssymb}
\newcommand{\be}{\begin{equation}}
\newcommand{\ee}{\end{equation}}
\newcommand{\bea}{\begin{eqnarray}}
\newcommand{\eea}{\end{eqnarray}}

\begin{document}

\setlength{\unitlength}{1mm}

\title{The Cosmology - Particle Physics Connection}

\author{Mark Trodden\footnote{trodden@physics.syr.edu}} 

\affiliation{Department of Physics, Syracuse University, 
Syracuse, NY 13244-1130, USA.}

\begin{abstract} 
Modern cosmology poses deep and unavoidable questions for fundamental physics. In this plenary talk, delivered in slightly different forms at the {\it Particles and Nuclei International Conference} (PANIC05) in Santa Fe, in October 2005, and at the {\it CMB and Physics of the Early Universe International Conference}, on the island of Ischia, Italy, in April 2006, I discuss the
broad connections between cosmology and particle physics, focusing on physics at the TeV scale, accessible at the next and future generations of colliders
\end{abstract}

\maketitle

\section{Introduction}
\label{intro}
I will begin by inventorying the energy budget of the universe, and pointing out the places where our
understanding is seriously hampered by issues that are firmly rooted in particle physics. I will then
go on to describe in broad terms the current status of our approaches to these issues. In some cases,
most notably dark matter and baryogenesis, collider physics may rule out or provide evidence for existing
proposals. On the other hand, if this is not the case, then precision measurements of physics at the
TeV scale may very well point the way to a new understanding of these fundamental cosmological
conundrums.

Beyond these topics, I will briefly speculate on possible connections between collider experiments
and one of the most esoteric cosmological concepts - dark energy.

Since this is a summary of a conference talk, my referencing will be very sparse, restricted to a few experimental results, some references to act as a caution about interpreting the acceleration data too literally, and some review articles from which the reader can find more complete references. I apologize in advance to any colleagues who may feel slighted by this decision.


\section{The New Cosmological Paradigm}
The data-driven revolution in cosmology cannot have escaped the notice of particle physicists. During the
last decade a host of new precision measurements of the universe have provided a clear and surprising
accounting of the energy budget of the universe. There now exists compelling evidence, from multiple
techniques, that the universe is composed of $5\%$ baryonic matter, $25\%$ dark matter and a whopping
$70\%$ dark energy, with negative pressure, sufficiently negative to cause the expansion of the 
universe to accelerate (although see~\cite{Deffayet:2001pu,Freese:2002sq,Dvali:2003rk,Carroll:2003wy,Capozziello:2003tk,Carroll:2004de} for some alternative views).

The best known evidence for this comes from two sources. 
The first is from Type Ia supernovae studies 
\cite{Riess:1998cb,Perlmutter:1998np}.
These data are much better fit by a universe dominated by a cosmological
constant than by a flat matter-dominated model.  This
result alone allows a substantial range of possible
values of $\Omega_{\rm M}$ and $\Omega_\Lambda$. However, if we independently
constrain $\Omega_{\rm M}\sim 0.3$, we obtain $\Omega_\Lambda \sim 0.7$, corresponding 
to a vacuum energy density 
$\rho_\Lambda \sim 10^{-8} {\rm ~erg/cm^3} \sim (10^{-3}{\rm ~eV})^4$.

The second is from studies of the small anisotropies in the Cosmic Microwave Background
Radiation (CMB), culminating in the WMAP satellite~\cite{hinshaw}. 
One very important piece of data that the CMB fluctuations give us is the value of $\Omega_{\rm total}$.  For a flat universe ($k=0$, $\Omega_{\rm total} =1$) we expect a peak in the power spectrum at 
$l\simeq 220$. Such a peak is seen in the WMAP data, yielding 
$0.98 \leq \Omega_{\rm total} \leq 1.08$ ($95\%$ c.l.) -- strong evidence for a flat universe.


\section{The Baryon Asymmetry of the Universe}
One would think that the baryonic component of the universe was well understood; after all, we are made
of baryons. However, from the point of view of cosmology, there is one fundamental issue to be understood. 

Direct observation shows that the universe around us contains no appreciable primordial antimatter. In addition, the stunning success of big bang nucleosynthesis rests on the requirement that, defining $n_{b({\bar b})}$ to be the number density of (anti)-baryons and $s$ to be the entropy density,
\begin{equation}
2.6\times 10^{-10} < \eta\equiv \frac{n_b -n_{\bar b}}{s} < 6.2\times 10^{-10} \ .
\end{equation}
This number has been independently determined to be $\eta =  6.1\times 10^{-10}\ ^{+0.3\times 10^{-10}}_{-0.2\times 10^{-10}}$ from precise measurements of the relative heights of the first two microwave background (CMB) acoustic peaks by the WMAP satellite.
Thus the natural question arises; as the universe cooled from early times, at which one would expect 
equal amounts of matter and antimatter, to today, what processes, both particle physics and 
cosmological, were responsible for the generation of this very specific baryon asymmetry? (For a review and references see~\cite{Trodden:1998ym,Riotto:1999yt}.)

If we're going to use a particle physics model to generate the baryon asymmetry
of the universe (BAU), what properties must the theory possess? This question
was first addressed by Sakharov in 1967, resulting in the
following criteria

\begin{itemize}
\item Violation of the baryon number ($B$) symmetry.
\item Violation of the discrete symmetries $C$ (charge conjugation)
      and $CP$ (the composition of parity and $C$)
\item A departure from thermal equilibrium.
\end{itemize}

There are {\it many} ways to achieve these. One particularly simple example is given by
Grand Unified theories (GUTs).
However, while GUT baryogenesis is attractive, it is not likely that the physics 
involved will be directly testable in the foreseeable future.

In recent years, perhaps the most widely studied scenario for generating 
the baryon number of the universe has been electroweak baryogenesis and I will
focus on this here.
In the standard electroweak theory baryon number is an exact global symmetry.
However, baryon number is violated
at the quantum level through nonperturbative processes. These effects are
closely related to the nontrivial vacuum structure of the electroweak theory.

At zero temperature, baryon number violating events are exponentially suppressed.
However, at temperatures above or comparable to the critical temperature
$T=T_c$ of the electroweak phase transition, $B$-violating vacuum transitions
may occur frequently due to thermal activation.

Fermions in the electroweak theory are chirally coupled to the gauge fields. 
In terms of the discrete symmetries of the theory,
these chiral couplings result in the electroweak theory being maximally
C-violating.
However, the issue of CP-violation is more complex.

CP is known not to be an exact symmetry
of the weak interactions, and is observed experimentally in the neutral 
Kaon system through $K_0$, ${\bar K}_0$ mixing. 
However, the relevant effects are parametrized by
a dimensionless constant which is no larger than $10^{-20}$. This appears
to be much too small to account for the observed BAU and so it is usual to turn
to extensions of the minimal theory. In particular the minimal supersymmetric standard
model (MSSM).

The question of the order of the electroweak phase transition is central to
electroweak baryogenesis. Since the equilibrium description of particle 
phenomena is extremely accurate at electroweak temperatures, baryogenesis 
cannot occur at such low scales without the aid of phase transitions.

For a continuous transition, the associated departure from
equilibrium is insufficient to lead to relevant baryon number production. 
For a first order transition quantum tunneling
occurs around $T=T_c$ and nucleation of bubbles of the true vacuum 
in the sea of false begins. At a particular temperature below $T_c$, bubbles
just large enough to grow nucleate. These are termed {\it critical} bubbles,
and they expand, eventually filling all of space and completing the transition.
As the bubble walls pass each point in space, the order
parameter changes rapidly, as do the other fields and this leads to a
significant departure from thermal equilibrium. Thus, if the phase 
transition is strongly enough first order it is possible to satisfy
the third Sakharov criterion in this way.

There is a further criterion to be satisfied. As the wall passes a
point in space, the Higgs fields evolve rapidly and the Higgs VEV changes from
$\langle\phi\rangle=0$ in the unbroken phase to $\langle\phi\rangle=v(T_c)$, the value
of the order parameter at the symmetry breaking global minimum of the finite 
temperature effective potential, in the broken phase. 
Now, CP violation and the departure from equilibrium occur while the Higgs field 
is changing. Afterwards, the point is
in the true vacuum, baryogenesis has ended, and baryon number violation
is exponentially supressed. Since baryogenesis is now over, 
it is
imperative that baryon number violation be negligible at this temperature in
the broken phase, otherwise any baryonic excess generated will be
equilibrated to zero. Such an effect is known as {\it washout} of the 
asymmetry and the criterion for this not to happen may be written as
\be
\frac{v(T_c)}{T_c} \geq 1 \ .
\label{washout}
\ee
It is necessary that this criterion be satisfied for any electroweak 
baryogenesis scenario to be successful.

In the minimal standard model, in which experiments now constrain the Higgs mass 
to be $m_H > 114.4$ GeV, it is clear from numerical simulations
that (\ref{washout}) is not satisfied. This is therefore a second
reason to turn to extensions of the minimal model.

One important example of a theory beyond the standard model in which these requirements can
be met is the MSSM.
In the MSSM there are two Higgs fields, $\Phi_1$ and $\Phi_2$. At one loop, a CP-violating
interaction between these fields is induced through supersymmetry
breaking. Alternatively, there also exists extra CP-violation through
CKM-like effects in the chargino mixing matrix. Thus, there seems to be
sufficient CP violation for baryogenesis to succeed.

Now, the two Higgs fields combine to give one lightest scalar Higgs $h$. 
In addition, there are also light {\it stops} ${\tilde t}$ (the
superpartners of the top quark) in the theory. These light scalar
particles can lead to a strongly first order phase transition if the 
scalars have masses in the correct region of parameter space. A detailed
two loop calculation~\cite{Carena:2002ss} and lattice results indicate that the allowed region is given by
\bea
m_h & \leq 120 {\rm GeV} \\
m_{\tilde t} & \leq m_t \ ,
\label{MSSMconstraints}
\eea  
for $\tan\beta \equiv \langle \Phi_2 \rangle/\langle \Phi_1 \rangle > 5$.
In the next few years, experiments at the Tevatron and the LHC should probe this range of Higgs 
masses and we should know if the MSSM is at least a good candidate for electroweak
baryogenesis.

What would it take to have confidence that electroweak baryogenesis within a particular SUSY
model actually occurred? First, there are some general predictions:
If the Higgs is found, the next
test will come from the search for the lightest stop at the Tevatron
collider. Important supporting evidence will
come from CP-violating effects which may be observable in $B$ physics.
For these, the preferred parameter space leads
to values of the branching ratio ${\rm BR}(b\rightarrow s\gamma)$
different from the Standard Model case. Although the exact value
of this branching ratio depends strongly on the value of the $\mu$
and $A_t$ parameters, the typical difference with respect to the
Standard Model prediction is of the order of the present experimental
sensitivity and hence in principle testable at 
the BaBar, Belle and BTeV experiments.

However, what is really necessary is to establish a believable model. For this we require precision 
measurements of the spectrum, masses, couplings and branching ratios to compare with theoretical
requirements for a sufficient BAU. Such a convincing case would require both the LHC and ultimately
the ILC if this is truly how nature works.


\section{Dark Matter}
Theorists have developed many different models for dark matter, some of which are accessible to
terrestrial experiments and some of which are not. There is not space to review all of these here. Rather, I
will focus on a specific example that is of interest to collider physicists (for a review and references see~\cite{Feng:2003zu}).

A prime class of dark matter candidates are Weakly Interacting Massive Particles (WIMPs). Such a
particle would be a new stable particle $\chi$. The evolution of the number density of these particles
in an expanding universe is
\begin{equation}
{\dot n_{\chi}}=-3Hn_{\chi}-\langle\sigma v\rangle(n_{\chi}^2-n_{eq}^2) \ ,
\end{equation}
where a dot denotes a time derivative, H is the Hubble constant, $\sigma$ is the annihilation 
cross-section and $n_{eq}$ is the equilibrium value of $n_{\chi}$.

In the early universe, at high temperature, the last term in this equation dominates and one finds the
equilibrium number density of $\chi$ particles. If this were always the case then today we would find 
negligible numbers of them and their energy density would certainly be too little to account for the dark
matter. However, as the universe expands it reaches a temperature, known as the 
{\it freeze-out temperature}, at which the evolution equation become dominated by the first term on the right-
hand side - the damping due to the the Hubble expansion. After this point, annihilations cease and the
distribution of $\chi$ particles at that time is merely diluted by the expansion at all later times, leading 
to an abundance that is much higher than the equilibrium one at those temperatures. 

In fact, to
a first approximation, the dark matter abundance remaining today is given by
\begin{equation}
\Omega_{DM}\sim 0.1 \left(\frac{\sigma_{\rm weak}}{\sigma}\right) \ ,
\end{equation}
where $\sigma_{\rm weak}$ is the typical weak interaction cross-section. From this one can clearly see
why it is that WIMPs get their name - weakly interacting particles yield the correct order of magnitude
to explain the dark matter.

What I have just described is a generic picture of what happens to a WIMP. Obviously, a specific 
candidate undergoes very specific interactions and a detailed calculation is required to yield the 
correct relic abundance. The most popular candidate of this type arises in supersymmetric extensions of the
standard model. Supersymmetry, of course, is attractive for entirely independent particle physics reasons.
However, a natural prediction of SUSY with low-energy SUSY breaking and R-parity is the existence of 
the lightest superpartner of the standard model particles. This Lightest Supersymmetric Particle (LSP)
is typically neutral, weakly interacting, with a weak scale mass, and hence can be a compelling dark matter
candidate. 

Weak scale SUSY has a large number of parameters. A detailed analysis requires us to focus
on particular models. It is common to use a model - minimal supergravity (mSUGRA) - 
described by just 5 parameters, the most important of which are the universal scalar mass
$m_0$ and the universal gaugino mass $M_{1/2}$, both defined at the
scale $M_{\rm GUT} \simeq 2 \times 10^{16}$GeV. In this framework the LSP is typically the
the lightest neutralino $\chi$ or the right-handed stau
$\tilde{\tau}_R$.  If it is a neutralino,
it is almost purely Bino over a large region of
parameter space, with a reasonable Higgsino component for $m_0
\geq 1$TeV.

It is, of course, very important to go beyond mSUGRA to understand all the possible ways for an LSP to
be the dark matter. However, mSUGRA does provide a crucial and manageable set of common models.

If SUSY is discovered at colliders, one 
would like to determine the relic density of such a
particle to an accuracy of a few percent, in order to compare with 
the known dark matter abundance. This requires a precise determination of the masses and couplings 
in the theory, a goal that, although challenging, may well be possible with the LHC and a linear collider.


\section{Dark Energy}
As I have mentioned, it is hard to see how one might make measurements directly relevant to the dark
energy problem in colliders. Nevertheless, in the interest of not giving up hope, and because we appear
to be extremely ignorant about this problem, I would like to mention at least one connection between the
cosmological constant, a candidate for the dark energy, and collider physics.

In classical general relativity the 
cosmological constant $\Lambda$ is 
a completely free parameter.
However, if we integrate over the quantum fluctuations of all modes of a quantum field
in the vacuum, we obtain a natural expectation for its scale. Unfortunately this integral
diverges, yielding an infinite answer for the vacuum energy.
Since we do not trust our understanding of physics at extremely high energies, 
we could introduce a cutoff energy, above which ignore any potential contributions, expecting 
that a more complete theory will justify this.  If the cutoff is at the Planck scale,
we obtain an estimate for the energy density in this component
\be
  \rho_{\rm vac} \sim M_{\rm P}^4 
  \sim (10^{18}{\rm ~GeV})^4 \ .
  \label{rhononsusy}
\ee

Unfortunately, a cosmological constant of the right order of magnitude to explain cosmic acceleration
must satisfy
\be
  \rho_{\rm vac} \sim (10^{-3}{\rm eV})^4 \ , 
  \label{rhoobs}
\ee
which is 120 orders of magnitude smaller than the above naive expectation.

A second puzzle, the {\it coincidence problem} arises because
our best-fit universe contains vacuum
and matter densities of the same order of magnitude. Since 
the ratio of these quantities changes rapidly as the universe expands.
there is only a brief epoch of the universe's history
during which we could observe the transition from domination by
one type of component to another. 

To date, I think it is fair to say that there are no approaches to
the cosmological constant problem that are both well-developed and compelling
(for reviews see~\cite{Carroll:2000fy,Peebles:2002gy,Sahni:1999gb}).
In addition, given the absurdly small mass scales involved, it is generally 
thought unlikely that collider physics will have any impact on this problem.
While I think this is probably true, I would like to emphasize a particular
connection between collider experiments and this problem.

As I have mentioned, a prime motivation for the next generation of accelerators is the possibility 
that supersymmetry might be discovered. At the risk of insulting some of my colleagues, when
one is constantly dealing with supersymmetric theories in the context of collider
signatures, it is easy to forget that supersymmetry is much more than a symmetry implying
a certain spectrum and specific relationships between couplings and masses. Supersymmetry is,
of course, a {\it space-time} symmetry, relating internal symmetry transformations with those 
of the Poincar\'{e} group. There is a direct connection between this fact and the vacuum energy.

The power of supersymmetry is that for each fermionic degree of freedom 
there is a matching bosonic degree of freedom, and vice-versa, so that their 
contributions to quadratic divergences cancel, allowing a resolution of the hierarchy problem.
A similar effect occurs when calculating the vacuum energy:
while bosonic
fields contribute a positive vacuum energy, for fermions the
contribution is negative.  Hence, if degrees of freedom exactly
match, the net vacuum energy sums to zero.  

We do not, however, live in a supersymmetric state (for example, there is no
selectron with the same mass and charge as an electron, or we would
have noticed it long ago).  Therefore, if supersymmetry exists, it must
be broken at some scale $M_{\rm SUSY}$.  In a theory with broken supersymmetry,
the vacuum energy is not expected to vanish, but to be of order
\be
  \rho_{\rm vac} \sim M_{\rm SUSY}^4 \sim (10^3{\rm ~GeV})^4 \ ,
\ee 
where I have assumed that supersymmetry is relevant to the hierarchy
problem and hence that the superpartners are close to experimental bounds.
However, this is still 60 orders of magnitude away from the observed value.

It is a crucial aspect of the dark energy problem to discover 
why it is that we do not observe a cosmological constant anything like this order of magnitude.
If we find SUSY at colliders and understand how it is broken, this
may provide much needed insight into how this occurs and perhaps provide new information
about the vacuum energy problem.


\section{Conclusions}
In this lecture I have tried to argue that particle physics and 
cosmology, as disciplines independent of one another, no longer exist; that our most fundamental
questions are the same and that we are approaching them in complementary ways. I have 
emphasized the deep connections between results obtained in existing colliders and expected from 
future ones and the puzzles facing cosmology regarding the energy budget of the universe.

From the familiar baryonic matter, through the elusive dark matter and perhaps all the way to the 
mysterious dark energy, collider experiments are crucial if we are to construct a coherent story of cosmic history. In conjunction with observational cosmology such experiments hold the key to unlock the deepest
secrets of the universe.

\acknowledgments
I would like to thank the organizers for two wonderful conferences. I also owe thanks to my co-members of the ALCPG Working Group on Cosmological Connections and in
particular my co-editors Marco Battaglia, Jonathan Feng, Norman Graf and Michael Peskin for
many useful conversations. This work was supported in part by the 
NSF under grant PHY-0354990. MT is a Cottrell Scholar of Research Corporation.

\end{document}